# SIMPLE FIT OF DATA RELATING SUPERMASSIVE BLACK HOLE MASS TO GALAXY PITCH ANGLE

Harry I. Ringermacher
General Electric Global Research Center, Schenectady, NY 12309

AND

Lawrence R. Mead
Dept. of Physics and Astronomy, University of Southern Mississippi, Hattiesburg, MS 39406

## ABSTRACT

Seigar, et al, have recently demonstrated a new, tight correlation between galactic central supermassive black hole (BH) mass and the pitch angle of the spiral arm in disc galaxies which they attribute to other indirect correlations. They fit a double power law, governed by five parameters, to the BH mass as a function of pitch. Noting the features of their fitted curve, we show that a simple linear proportion of the BH mass to the cotangent of the pitch angle can obtain the same fit, within error. Such a direct, elegant fit may help shed light on the nature of the correlation.

*Subject Headings:* galaxies: fundamental parameters – galaxies: spiral – galaxies: structure

Seigar, et al.(2008) demonstrate a compelling correlation between the central BH mass of a spiral galaxy and the angle of pitch of its arms – assuming logarithmic spirals (constant pitch). The spiral pitch angle P is the angular difference between the spiral tangent at point $(r,\phi)$ and a circle tangent at that point. It is defined from (Binney 1987):

$$cot(P) = r(\phi)\frac{d\phi}{dr} \qquad (1)$$

Seigar, et al. make it clear that they have inferred this correlation indirectly from other observed correlations. They present two curves– a convenient log-linear fit over a smaller central pitch domain and the more general "double-power-law model", based on the "Nuker" law (Lauer, 1995), that covers both large and small pitch. The so-called Nuker law (Eqn. (3) below) was intended to describe mass density distributions and the pitch angle, P, would ordinarily be replaced by a radius, r. The law can simultaneously describe distinctly different small radius and large radius behavior. The resulting form could, for example, describe either a cuspy core or a constant core mass density, depending on choice of parameters. In the present case, it appears to be simply an excellent dual power law fitting equation for differing large and small pitch behavior.

Seigar states that there appears to be an upper pitch cutoff at about 42º for very small BH mass and may indicate a minimal mass. At the same time he states that his fit is not applicable below pitch angles of ~7º – applying here as well. These fits are described by their equation (1) – including erratum ( Seigar et al. 2008 Erratum) –

$$\log_{10} M_{BH} = (8.44 \pm 0.10) - (0.076 \pm 0.005)P \qquad (2)$$

and
$$M_{BH} = 2^{(\beta-\gamma)/\alpha} M_{BH_b} \left(\frac{P_b}{P}\right)^{\gamma} \left[1 + \left(\frac{P}{P_b}\right)^{\alpha}\right]^{(\gamma-\beta)/\alpha} \qquad (3)$$

with best fit parameters; $\alpha = 23.5, \beta = 126.1, \gamma = 2.92, M_{BH_b} = 1.72 \times 10^4 M_\odot, P_b = 40.8°$.

In Figure 1 we present a simpler fit to the same data. The fit is given by:

$$M_{BH} = M_0 \left( \cot(P - P_0) - I \right) \qquad (4)$$

This fit was motivated by the observation that the Seigar fit closely resembles a cotangent function. The three parameters are $M_0 = 2 \times 10^6 M_\odot$, $P_0 = 6.446°$ and $I = 1.3881$. We have also included equation (2) on the graph as a reference. The very definition of pitch angle is through its cotangent, equation (1), so the fit of equation (4) may be of particular interest. Our plot of equation (4) is nearly identical to Seigar's within his error. We have overlaid the two equations in Fig.1 for convenience. In obtaining this fit, due to lack of data on the very low pitch side, the parameters $M_0$ and $I$ were selected to best match the Seigar curve end points and constrained while a best fit value was found for $P_0$, the most sensitive parameter. Our curve was end-adjusted to precisely pass through Seigar's lowest pitch point (Seigar, communication), 7.1°, since that was M31 data and presumably quite accurate. Seigar's cutoffs at low and high pitch thus appear naturally in our fit since the cotangent has two singularities. The low pitch end is singular at $P = P_0 \simeq 6.4°$ while the high pitch end is singular at about 42°. More data at the low pitch end would better constrain the curve and the singular point could be adjusted to as low as 4°-5°, but not to 0° (and still fit the remaining data). Seigar's curve, on the other hand, is not singular at either end so there is a significant qualitative difference between our two fits. Clearly, the BH mass tends rapidly to zero with an upper pitch of ~ 42°. Thus our curve would fit the scenario that spiral galaxies with pitch greater than about 42° either do not have central supermassive massive black holes at all – suggesting a BH mass cutoff - or there is a maximum pitch as BH mass tends to zero.

The presence of the cutoffs may express some physical reality. Whether future low pitch data are consistent with singularities remains to be seen. The high pitch data is already consistent with a singularity. Thus if incoming low pitch data fit one or the other of the curves, that should provide some insight as to the nature of the correlation. For example, at low pitch, a rapidly rising BH mass would favor equation (4) and suggest a low pitch cutoff, which should stimulate further theoretical investigation for the correlation.

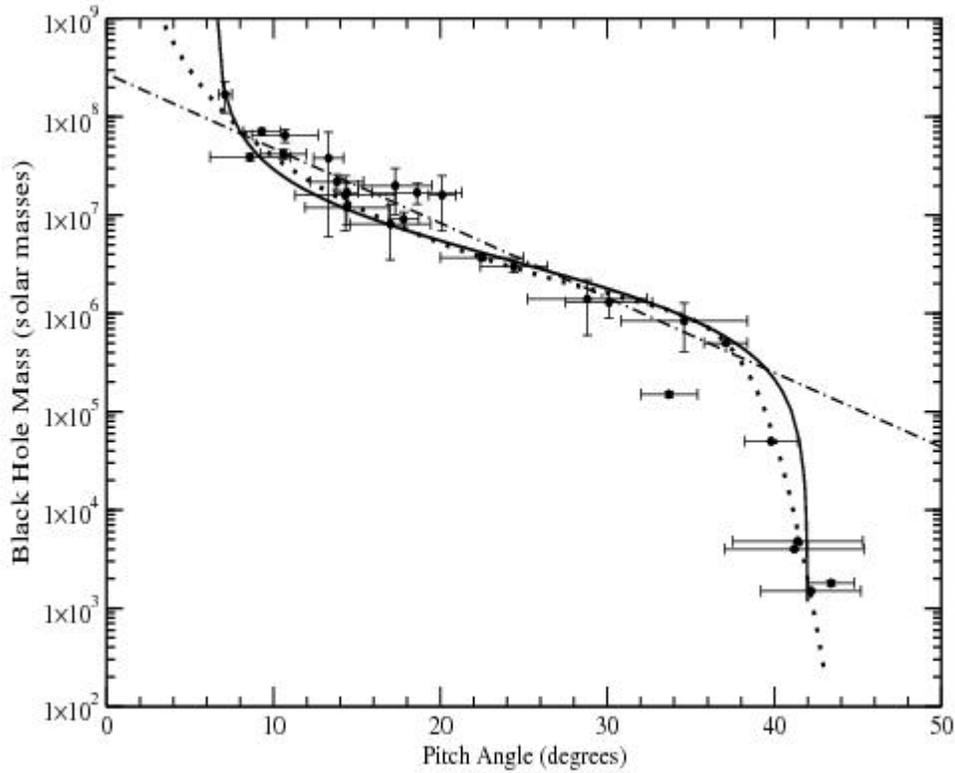

Fig. 1. – Plot of Black hole mass as a function of spiral arm pitch angle from eq. (4) (dark curve). Seigar's log-linear fit (eq. (1) – dadot) and "Nuker" fit ,(eq. (3)-dots) have been included for reference.

ACKNOWLEDGEMENTS

We are grateful to Marc Seigar for his helpful comments.